\documentclass[twocolumn, aps, longbibliography, nofootinbib]{revtex4-2}
\usepackage{float}          
\usepackage{algorithm}
\usepackage{algpseudocode}
\usepackage{graphicx}
\usepackage{dcolumn}
\usepackage{bm}
\usepackage{amsmath}
\usepackage{mathtools}
\usepackage{comment}
\usepackage{appendix}
\usepackage{tikz}
\usepackage{chemformula}
\usepackage[colorlinks=true]{hyperref}
\usepackage{subfigure}
\usepackage{changepage}
\usepackage{amssymb}
\usepackage{graphicx}
\usepackage{textcomp}
\usepackage{xcolor}
\usepackage{hyperref}

\usepackage{ulem} 

\newcommand{\BLUE}[1]{\textcolor{black}{#1}}

\definecolor{darkgreen}{RGB}{0,100,0}

\newcommand{\bra}[1]{\langle #1 \rvert}
\newcommand{\ket}[1]{\lvert #1 \rangle}

\begin{document}

\preprint{APS/123-QED}

\title{How to \textit{really} measure operator gradients in ADAPT-VQE}

\author{Panagiotis G. Anastasiou$^{1,3}$}\email{panastasiou@vt.edu}
\author{Nicholas J. Mayhall$^{2,3}$}
\author{Edwin Barnes$^{1,3}$}
\author{Sophia E. Economou$^{1,3}$}\email{economou@vt.edu}

\affiliation{$^1$Department of Physics, Virginia Tech, Blacksburg, VA 24061, USA \\
	$^2$Department of Chemistry, Virginia Tech, Blacksburg, VA 24061, USA
 \\
 $^3$Virginia Tech Center for Quantum Information Science and Engineering, Blacksburg, VA 24061, USA}

\date{\today}

\begin{abstract}
ADAPT-VQE is a leading variational quantum algorithm as it circumvents the choice-of-ansatz conundrum by iteratively growing compact and arbitrarily accurate problem-tailored ans\"atze. However, for molecular Hamiltonians and hardware-efficient operator pools, the gradient-measurement step of the algorithm requires the estimation of $\mathcal{O}(N^8)$ observables, which may represent a bottleneck for relevant system sizes on real devices. \BLUE{We present an efficient strategy for measuring the gradients of the three best-performing hardware-efficient operator pools by partitioning the required observables into $\mathcal{O}(N^3)$-sized sets of commuting Paulis that can be simultaneously measured with an at most $N-3$ CNOT overhead.} We argue that our approach is robust to shot-noise effects and show that measuring the pool gradients is, in fact, \BLUE{not $\mathcal{O}(N^4)$, but only about $4N$ times as expensive as measuring the energy once}. Our proposed measurement strategy significantly ameliorates the measurement overhead of ADAPT-VQE and brings us one step closer to practical implementations on real devices. 

\end{abstract}

\maketitle


\section{\label{sec:intro}Introduction}

The Variational Quantum Eigensolver (VQE) has garnered significant attention in the fields of quantum simulation and chemistry \cite{QCC_review, QCQC_review, QCC_review_recent}, as one of the most promising algorithms for attaining useful quantum advantage on noisy intermediate-scale quantum (NISQ) devices. It is a hybrid quantum-classical algorithm designed to find the ground state and lowest eigenvalue of a given Hamiltonian based on the variational principle of quantum mechanics \cite{Peruzzo2014OGVQE}. In the VQE, an upfront-chosen parametrized guess wavefunction (ansatz) is prepared on the quantum processor, the expectation value of the Hamiltonian with respect to the ansatz is measured, and using some optimization scheme running on a classical machine the parameters are updated until a preset convergence criterion is met. Since the quantum processor is only used for state preparation and energy measurement, the deep evolution circuits required for other eigenvalue-finding algorithms \cite{kitaev1995PEA} are avoided, albeit at the expense of additional state preparations. We note that much of the theory of VQE has been extended to cost functions other than the energy \cite{zhang2020varianceVQE}, target states besides the ground \cite{Izmaylov_constrainedVQE}, and problems beyond quantum simulation \cite{qaoa}.

Paramount to the success of a VQE experiment is the choice of ansatz, which should be expressive enough so that it gives a reasonable upper bound for the ground state energy, sufficiently compact so that it can be successfully prepared on a NISQ device, and have a parameter landscape conducive to classical optimization. To this day, most proposed ans\"atze are roughly divided into two families: the hardware-efficient, and the chemically-inspired ones. The former are relatively easy to implement on quantum devices as they consist of layers of single-qubit rotations and native entangling gates \cite{Kandala2017}. However their structure is completely agnostic to the problem at hand, and to ensure expressivity they reach regions of the Hilbert space that lie beyond the solution. As a result, they are often plagued by barren plateaus: exponentially flat neighborhoods of the cost function landscape that hinder classical optimization \cite{holmes2022connecting}. The latter, on the other hand, consist of circuits implementing unitary transformations inspired from classical computational chemistry, with prototypical example the Unitary Coupled-Cluster (UCC) ansatz, and are known to converge reliably in simulations \cite{UCCrev}. Nonetheless, such ans\"atze are prepared by relatively deep circuits, and the performance of their low Trotter order approximations has been shown to be sensitive to the Trotterization ordering \cite{GrimsleyIllDefined}. 

To address these issues, Grimsley et al. introduced the Adaptive Derivative-Assembled Problem-Tailored ansatz (ADAPT)-VQE algorithm which circumvents the choice-of-ansatz conundrum by iteratively growing problem-specific ans\"atze on the fly \cite{OG_ADAPT}. Starting from some initial state, e.g., the Hartree-Fock solution and a user-defined operator pool, a parametrized unitary is appended to the ansatz in each iteration, followed by ordinary VQE. Assuming that the VQE subroutine terminates at a minimum, the gradient of the energy with respect to the variational parameter of each candidate operator from the pool is measured, and the operator with the largest gradient magnitude is added to the ansatz. The iterative procedure is repeated until the pool gradient norm falls below a chosen threshold, thus yielding in principle arbitrarily accurate ans\"atze. Furthermore, since a high-gradient operator is added to the guess in each iteration, the VQE procedure avoids barren plateaus and is resistant to local traps \cite{Grimsley2023}. However, the well-optimizable and NISQ-friendly ans\"atze come with significant measurement overhead associated with measuring the gradient of every pool operator in each ADAPT iteration. For electronic Hamiltonians with $\mathcal{O}(N^4)$ terms and \BLUE{the gate-efficient qubit~\cite{qubitADAPT}, qubit-excitation (QEB)~\cite{QEB_ADAPT_VQE}, and Coupled-Exchange Operator (CEO)~\cite{Ramoa2025} } pools whose size grows as $\mathcal{O}(N^4)$, evaluating the pool gradient requires measuring $\mathcal{O}(N^8)$ observables where $N$ is the number of qubits. 

The success of ADAPT-VQE in silico has prompted several theoretical efforts to ameliorate the steep scaling of the gradient-measurement step of the algorithm. In Refs.~\cite{qubitADAPT, shkolnikov2021avoiding}, it was shown that pools with as few as $2N-2$ carefully chosen operators can be complete: ADAPT can arrive at any final state in the Hilbert space if sufficiently many unitaries generated by the operators in these pools are added to the ansatz. Although such pools require measuring only $\mathcal{O}(N^5)$ observables, how their use affects the depth of the final ansatz and which $2N-2$ operators constitute an optimal minimal complete pool (MCP) are still open questions. Liu et al. \cite{RDM-ADAPT} introduced a gradient estimation scheme based on the approximate reconstruction of the three-body reduced density matrix (3-RDM) that avoids the measurement overhead altogether but results in longer ans\"atze, as well as a strategy for pre-screening pool operators which reduces the number of gradients to be measured in each iteration with a smaller trade-off in ansatz accuracy. In the same vein,  Nyk\"anen et al. introduced the Adaptive Informationally complete (IC) generalised Measurements (AIM)-ADAPT-VQE algorithm \cite{POVM-ADAPT} in which IC POVM data acquired for cost function estimation is reused to estimate the pool gradients. Although the results for the single 8-qubit system studied are promising, whether the algorithm scales to relevant system sizes is uncertain, especially considering that IC-POVMs in general consist of $4^N$ operators to be sampled. Moreover, Majland and coworkers claimed that it is possible to estimate the operator pool gradients classically by replacing the wavefunction with a distribution of Slater Determinants obtained from measurements in the computational basis \cite{FAST-VQE}. Since their heuristic gradient expression essentially replaces quantum amplitudes with frequencies, discarding all phase information, it is not expected to be accurate for any but the most weakly correlated systems. \BLUE{More recently, Ref. \cite{ikhtiarudin2025shotefficientadaptvqereusedpauli} proposed simultaneously measuring qubit-wise commuting terms of each ADAPT gradient while reusing energy estimation data achieving roughly a $60\%-70\%$ shot reduction, with greater reduction associated with the smaller systems simulated. Lastly, Huang and Izmaylov reformulated the operator selection problem into a Best Arm Identification one, and used Successive Elimination to weed out operators and consistently reduce the gradient measurement cost by $70\%-90\%$, observing once again that more iterations and total operators may be required to reach chemical accuracy due to occasional suboptimal selection \cite{Huang2026}.}

The aforementioned studies either restrict the pool sizes or employ approximations for the estimation of the pool operator gradients instead of measuring them directly, trading off ansatz compactness for measurement economy at various degrees. But is it possible to reduce the state preparation cost of ADAPT-VQE in a provably scalable manner without sacrificing hardware efficiency or accuracy? 

\begin{figure}
  \includegraphics[width=1\linewidth]{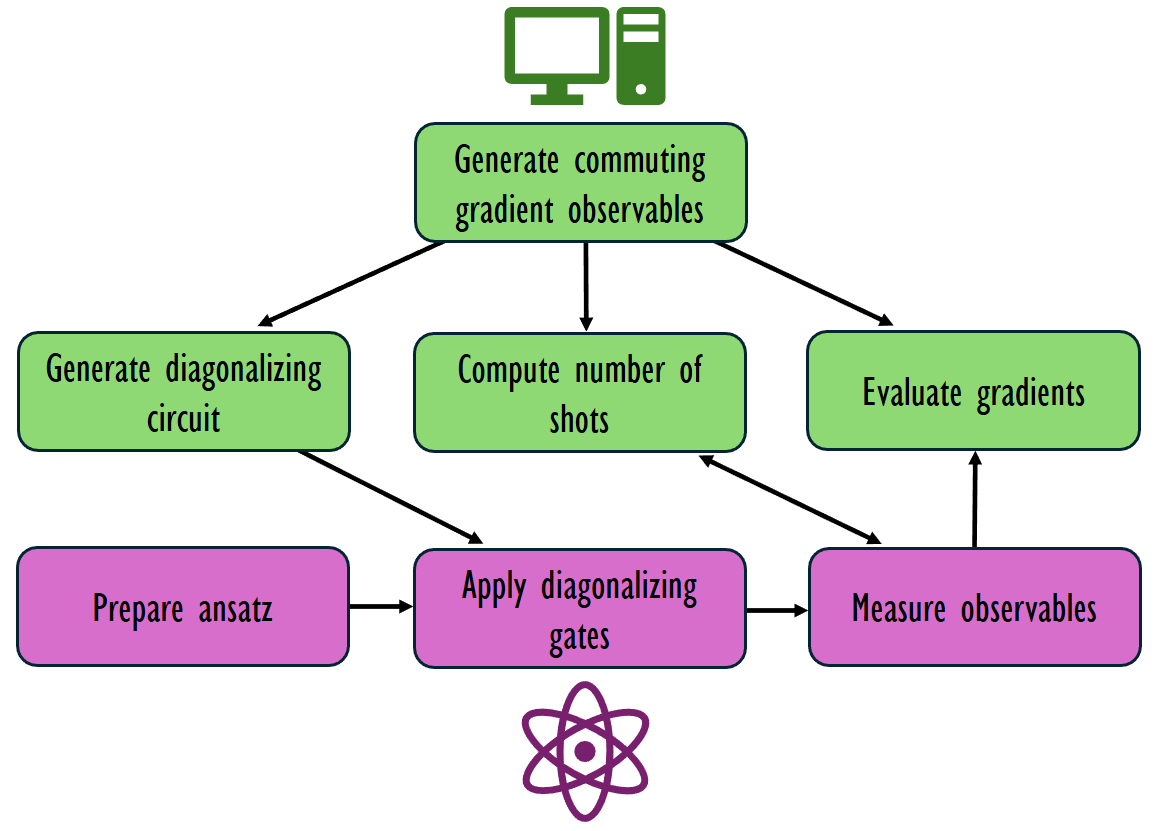}
  \caption{\BLUE{Our measurement procedure begins by generating the sets of commuting gradient observables as explained in Sec. \ref{subsec:singlepivot}, which are used to compute a quasi-optimal number of shots (Sec. \ref{subsec:shot_noise}) and to generate a diagonalizing circuit (Sec. \ref{subsec:diagonalizing_circ}) for each set. The ADAPT ansatz is prepared on the quantum computer, followed by a diagonalizing circuit and measurement. Measurement results are used to inform subsequent rounds of measurement (via variance estimation) and to evaluate the gradients of interest.}}
  \label{fig:summary}
\end{figure}

In this work\BLUE{, overviewed in FIG. \ref{fig:summary},} we present a way to arrange the $\mathcal{O}(N^8)$ observables \BLUE{necessary to evaluate the qubit~\cite{qubitADAPT}, QEB~\cite{QEB_ADAPT_VQE}, and CEO~\cite{Ramoa2025} pool gradients} into only $\mathcal{O}(N^5)$ mutually commuting sets. \BLUE{We prove that diagonalizing circuits for simultaneous measurement with at most $N-3$ CNOT gates exist and show how to construct them.} Additionally, we show that our proposed grouping strategy automatically takes into account the importance of each observable of interest, and offers a way to optimally allot state preparations to the different sets under certain assumptions. Lastly, we show that measuring the gradient of the entire pool is more expensive than a single VQE iteration by merely a factor \BLUE{of $4N$}. We expect that our approach will play a key role in implementing ADAPT-VQE on real devices for problems beyond toy examples.

The paper is organized as follows. In Sec.~\ref{sec:adapt} we briefly review the ADAPT-VQE algorithm and hardware-efficient pools. Our main results are given in Sec.~\ref{sec:results}, where we present a gradient measurement strategy for hardware-efficient pools that requires at most a linear measurement overhead compared to standard VQEs. \BLUE{Our theoretical results are backed by numerical experiments in Subsec. \ref{subsec:simulations}.} Secs.~\ref{sec:discussion} and \ref{sec:conclusions} contain further discussion of the results and conclusions.
 
\section{ADAPT-VQE details}\label{sec:adapt}


Although (akin to ordinary VQE) ADAPT is quite versatile \cite{gibbs-adapt, variance-adapt}, in this section we present it with the objective of finding the lowest eigenvalue of a given Hamiltonian, i.e., with cost function $E=\langle H \rangle$. The user begins by defining an operator pool $\mathcal{A}=\{A_i\}$, a collection of antihermitian generators whose parametrized exponentials are used to form the ansatz. In addition, a pool gradient norm threshold $g_{min}$ is set, and a reference state $\ket{\Psi_0}$ is chosen, usually the Hartree-Fock (HF) ground state for chemical problems. The user also provides the Pauli decomposition of the problem Hamiltonian, through, e.g., the Jordan-Wigner (JW) fermion-to-qubit mapping \cite{BKmapping}. \BLUE{The algorithm proceeds as follows:}

\begin{algorithm}[H] 
\caption{ADAPT-VQE}\label{alg:adapt-vqe-x} 
\begin{algorithmic} 
\State $k \gets 0$ 
\State $\mathcal{U}_0 \gets \mathbb{I}$ 
\State $\text{Converged} \gets \text{False}$ 
\While{$\text{Converged} = \text{False}$} 
    \For{$A_i \in \mathcal{A}$} 
        \State $g_i \gets \Bigl|\frac{\partial}{\partial \theta_i} \bra{\Psi_{0}}\mathcal{U}_k^\dagger (\boldsymbol{\theta}) e^{-\theta_iA_i}He^{\theta_iA_i} \mathcal{U}_k (\boldsymbol{\theta}) \ket{\Psi_{0}}\Bigr|_{\theta_i=0, \boldsymbol{\theta}}$
    \EndFor 
    \If{$\|\boldsymbol{g}\| < g_{\min}$} 
        \State $\text{Converged} \gets \text{True}$ 
    \EndIf 
    \State $j \gets \text{index of largest element of } \boldsymbol{g}$ 
    \State $k \gets k+1$ 
    \State $\mathcal{U}_k (\boldsymbol{\theta}) \gets e^{\theta_jA_j}\mathcal{U}_k (\boldsymbol{\theta})$ 
    \State $\boldsymbol{\theta} \gets \underset{\boldsymbol{\theta}}{\operatorname{argmin}} \bra{\Psi_{0}}\mathcal{U}_k^\dagger (\boldsymbol{\theta}) H \mathcal{U}_k (\boldsymbol{\theta})\ket{\Psi_{0}}$ 
    \State $E_k \gets \bra{\Psi_{0}}\mathcal{U}_k^\dagger (\boldsymbol{\theta}) H \mathcal{U}_k (\boldsymbol{\theta})\ket{\Psi_{0}}$ 
\EndWhile 
\State \Return $E_k, \boldsymbol{\theta}$ 
\end{algorithmic} 
\end{algorithm}

We note that adding an operator to the ansatz does not drain the pool: a single operator can be added to the ansatz multiple times, though not in consecutive iterations. In the $k$-th iteration, the ansatz has the form:

\begin{equation}\label{eq:ansatz_form}
    \ket{\Psi_k}=e^{\theta_k A_k}...e^{\theta_1 A_1}\ket{\Psi_0},
\end{equation}

and the energy gradient with respect to the variational parameter of candidate operator $A_i$, $\theta_i$ in the $(k+1)$-th iteration, using the antihermiticity of the pool operators becomes

\begin{align}\label{eq:gradient}
    \frac{\partial E}{\partial \theta_i}\bigg|_{\theta_i=0} &= \left[\frac{\partial}{\partial \theta_i} \bra{\Psi^{(k)}}e^{-\theta_iA_i}He^{\theta_iA_i}\ket{\Psi^{(k)}}\right]
    \bigg|_{\theta_i=0} \nonumber \\
    &= \bra{\Psi^{(k)}}[H,A_i]\ket{\Psi^{(k)}},
\end{align}  
noting for clarity that in Eq.~\eqref{eq:ansatz_form} the operator subscripts denote the order in which they were appended to the ansatz, whereas in Eq.~\eqref{eq:gradient} the subscript $i$ denotes a specific operator arranged in no particular order in the pool.

\BLUE{Our aim is to reduce the cost of estimating} $\langle[H,A_i]\rangle$ for every operator $A_i$ in pool $\mathcal{A}$, where $H$ can be written as a weighted sum of Pauli strings, and we take $A_i$ to be individual Pauli strings, perhaps with multiplicative phases absorbed:
\begin{align}\label{eq:gradient_commutator}
[H,A_i]=\bigg[\sum_j h_jH_j, A_i\bigg] 
\\ \nonumber   =\sum_jh_j[H_j,A_i]=\sum_j c_{ij}C_{ij},
\end{align}
where $h_j$, $c_{ij}$ (which either vanish or are proportional to $h_{j}$) are complex coefficients and $H_j$, $A_{i}$, and $C_{ij}$ are Pauli strings. \BLUE{In particular, we focus on the operator pool introduced by Tang et al.~\cite{qubitADAPT}, which includes $iY_iX_j$, $iY_iX_jX_kX_l$, $iX_iY_jY_kY_l$, and is dubbed ``qubit pool'', and point the interested reader to Appendix~\ref{appendix_operator_pools} for a brief review of hardware-efficient operator pools. We note that even though we focus on the qubit pool, our results directly apply to the QEB and CEO pools because their operators are linear combinations of mutually commuting qubit pool operators, and because both commutators and expectation values are linear.}

\section{Main results}\label{sec:results}

The na\"ive way to estimate the pool gradients would be to loop over operators $A_i$, evaluating $\sum_{j} c_{ij}C_{ij}$ and measuring the expectation values of the observables $C_{ij}$ for each pool operator sequentially.
 A somewhat more sophisticated approach to reduce the number of state preparations required for the task would be to generate the list of all $C_{ij}$ with nonvanishing $c_{ij}$, use one of the many available heuristics \cite{min_clique_Izmaylov, Crawford2021efficientquantum, gokhale2019minimizing} to group the resulting terms into sets of mutually commuting observables, and for each such set perform a rotation into a shared eigenbasis \cite{GC_toQWC_Unitaries_Izmaylov, Crawford2021efficientquantum, gokhale2019minimizing, vandenBerg2020circuitoptimization} followed by measurement. To the best of our knowledge all available heuristics for Pauli observable partitioning lack performance guarantees and have complexities $\mathcal{O}(Np^2)$ or higher, where $p$ is the number of Pauli observables. Applied to our task at hand where $p \sim N^8$, this would imply a formidable $\mathcal{O}(N^{17})$ classical preprocessing overhead. In this section, we give a recipe for collecting the commutators of all pool operators with a single Hamiltonian term into $2N$ sets for the hardware-efficient pools \BLUE{discussed in Appendix~\ref{appendix_operator_pools}, thus partitioning $\mathcal{O}(N^{8})$ terms into only $\mathcal{O}(N^{5})$ sets}.

\subsection{\label{subsec:simultaneous_commutation_proof}The commutators of a set of commuting Pauli words with any one Pauli word commute}
Our aim is to find an efficient partitioning of the commutators of many pool operators with a Hamiltonian, assuming that we have a decomposition of both as sums of Pauli words. As usual, by Pauli words (also known as Pauli strings) we mean tensor products of Pauli matrices (including the identity), i.e.,  $O=a_0\otimes a_1\otimes...$ where $a_i \in \{I,X,Y,Z\}$, and to avoid notational clutter we omit identities and the tensor product symbol and add qubit indices everywhere else, e.g., $X_1Y_2$ acting on the Hilbert space of 4 qubits should be read as $I\otimes X \otimes Y \otimes I$.
 Now consider the commutator of two commutators, $[[P,S_i],[P,S_j]]$ for Pauli
 words $P$, $S_i$, and $S_j$. If either of the inner commutators vanish, the
 outer one vanishes trivially. Considering only non-vanishing inner commutators
 and using the fact that Pauli words that do not commute anticommute, we have
\begin{align*}
    [[P,S_i],[P,S_j]]=4[PS_i,PS_j]
    =4(-S_iPPS_j+S_jPPS_i)\\=4[S_j,S_i].
\end{align*}
This implies that $[[P,S_i],[P,S_j]]$ vanishes if $S_i$ and $S_j$ commute, and by extension, if we define the set of operators $\{O_k=[P,S_k]|S_k\in\mathcal{S}\}$, where $P$ is any Pauli word, and $\mathcal{S}$ is any set of mutually commuting Pauli words, then $[O_k,O_\ell]=0$ $\forall k,\ell$. Since all commutators revolve around $P$, from this point onward we call $P$ the pivot of the set. Since we are interested in measuring the commutators of the Hamiltonian and each pool operator, we have the freedom to choose the pivots to be operators from our pool, or the individual terms in the Hamiltonian, \BLUE{with one of the two choices offering a significant advantage: Choosing the pivots to be pool operators would imply partitioning the $\mathcal{O}(N^{4})$ Hamiltonian terms into $\mathcal{O}(N^{3})$ sets, as discussed in Appendix~\ref{appendix_Hlinearpartitions}, while choosing the pivots to be Hamiltonian terms allows us to group the $\mathcal{O}(N^{4})$ pool operators into exactly $2N$ sets, as we show next.}

\subsection{\label{subsec:singlepivot} Hardware-efficient pool operators can be partitioned into $2N$ mutually commuting sets}

In this subsection, we show that combining qubit pool operators instead of Hamiltonian terms leads to only ${\cal O}(N^5)$ sets of mutually commuting observables instead of ${\cal O}(N^7)$. The simple form, lack of Pauli $Z$ strings, and uniform weights of qubit pool operators prompt us to move away from the usual heuristics used for Hamiltonian term grouping and seek a deterministic way to arrange them into commuting sets.

\BLUE{For any four distinct qubit indices $i,j,k,l$, the double-like qubit pool operators (which dominate in number for larger systems) are $iY_iX_jX_kX_l$ and $iX_iY_jY_kY_l$, for all permutations of indices, which give 8 unique operators. It is easy to see that each of these operators fails to commute with the rest on an even number of qubits, therefore all 8 commute. However, due to the non-transitive nature of commutation, if we group together some or all of these 8 pool operators into a single mutually commuting set, this precludes alternative, more efficient groupings; for example, combining $iY_iX_jX_kX_l$ with $iX_iX_jX_kY_l$, makes $iY_iX_jX_kX_{l+2}$ ineligible for addition to the set. This implies that in each commuting set we would have at most $\mathcal{O}(N)$ pool operators, and the number of sets would once again scale as $\mathcal{O}(N^3)$. Can we do better by using qubit-wise commutation?}
 
\BLUE{To differentiate the two types of operators, we will refer to them as $YXXX$ and $XYYY$ operators for brevity. Since the $YXXX$ operators have one $Y$ Pauli and 3 $X$ each, there is an obvious way to arrange all $\mathcal{O}(N^4)$ of them into only $N$ sets of qubit-wise commuting operators: create $N$ sets and in the $i^{\mathrm{th}}$ set put all $Y_iXXX$ operators together. In each set, we will call the fixed index the anchor. That is, we form a set $\{iY_0X_jX_kX_l\}$ with anchor qubit 0, another one $\{iY_1X_jX_kX_l\}$ with anchor qubit 1, etc., with indices $j, k, l$ varying freely. Similarly partitioning the $XYYY$ operators (anchored on the $X$ indices) implies that we can arrange \textit{all} $\mathcal{O}(N^4)$ qubit pool operators into only $2N$ commuting sets.} 

\BLUE{
Since the number of Hamiltonian terms scales as $\mathcal{O}(N^4)$, we can arrange all observables required for the evaluation of the commutators of the Hamiltonian with all $\mathcal{O}(N^4)$ qubit operators into only $\mathcal{O}(N^5)$ commuting sets by invoking the proof in Subsec.~\ref{subsec:simultaneous_commutation_proof}. Here, we take the pivot $P$ to be a term in the Hamiltonian, while $\mathcal{S}$ is one of the $2N$ sets of mutually commuting pool operators. The single-like operators can be readily accommodated into the double-like sets: e.g., the $iY_1X_3$ operator can enter both the $YXXX$ set anchored on qubit 1 and the $XYYY$ set anchored on qubit 3.
}

\BLUE{
QE and CEO pool gradients can be obtained by taking sums of qubit pool operators. From here onward we focus solely on the qubit pool and double-like operators. See Algorithm~\ref{alg:generate_observables} for a high-level recipe for generating the commuting sets for all qubit pool gradients and the first two columns of TABLE~\ref{table_circuit_1} for a realistic example with a single pivot.}

\subsection{\label{subsec:diagonalizing_circ}\BLUE{Diagonalizing circuits with at most $N-3$ CNOT gates}}

\BLUE{For the simultaneous measurement of gradient observables we require circuits that rotate the computational basis into a basis in which all ${\cal O} (N^3)$ observables in a commuting set are diagonal; that is, they can be simultaneously measured by single-qubit measurements along the $Z$-axis and classical post-processing. 
 In this subsection, we show that for the qubit pool there exist diagonalizing circuits that contain at most $N-3$ CNOT gates, we illustrate a general procedure for finding them, and we apply this procedure step-by-step to a simple but illustrating 8-qubit example.
}

\BLUE{
We begin by considering only particle number parity- and $S_z$ parity-preserving weight-4 qubit pool operators~\footnote{\BLUE{that is, each one acts on an even number of even qubits and an even number of odd qubits where even qubits correspond to $\alpha$ spin-orbitals and odd qubits to $\beta$ spin-orbitals}} of the form $iY_jX_aX_bX_c$, where $j$ is fixed. For concreteness, consider $j=0$. We consider the worst case in which all of these operators anticommute with $P$, in which case we have $(N/2-1)\binom{N/2}{2}+\binom{N/2-1}{3}$ commuting operators $PY_0X_aX_bX_c$ to simultaneously measure in order to obtain their contributions to the gradients. We can multiply all these operators by one of them, say $PY_0X_1X_2X_3$, to arrive at a new, equivalent set of operators that includes $PY_0X_1X_2X_3$ and $(N/2-1)\binom{N/2}{2}+\binom{N/2-1}{3}-1$ stabilizers that contain only identities and $X$'s. These $X$-strings are generated by the set $\{X_iX_{i+2}|i=1,\ldots,N-3\}$, since they and their products have support on an even number of even qubits (not including 0) and on an even number of odd qubits. Because $\mathrm{CNOT}_{i,i+2}X_iX_{i+2}\mathrm{CNOT}_{i,i+2}=X_i$, we can convert the set $\{X_iX_{i+2}|i=1,\ldots,N-3\}$ into $\{X_i|i=1,\ldots,N-3\}$ using $N-3$ CNOT gates. These CNOT gates necessarily transform $PY_0X_1X_2X_3$ into a Pauli string of the form $A(\prod_{i=1}^{N-3}B_i)CD$, where the $B_i$ are either $I$ or $X$, and $A$, $C$, $D$ can be any single-qubit Paulis, since this operator must commute with the set $\{X_i|i=1,\ldots,N-3\}$. At this point, all the generators have only identities or $X$'s on qubits $1,\ldots,N-3$, and these can be converted to identities and $Z$'s using Hadamard gates, while the Pauli operators $A$, $C$, $D$ on qubits 0, $N-2$, $N-1$ can also be converted to identity or $Z$ using single-qubit Clifford gates. Thus, at most $N-3$ CNOT gates are needed to simultaneously diagonalize the mutually commuting set of operators $PY_0X_aX_bX_c$. If some of the operators $Y_0X_aX_bX_c$ commute with $P$, then it may be possible to diagonalize the remaining operators using fewer CNOTs. The same analysis clearly holds for any other value of $j$ and for operator sets of the form $PX_jY_aY_bY_c$.}

\BLUE{We illustrate this approach with an 8-qubit example. We choose the Hamiltonian term $P=Y_1Z_2Z_3Z_4Z_5Z_6Y_7$ and set qubit $0$ as the anchor qubit, in which case all operators of the form $iY_0X_aX_bX_c$ fail to commute with $P$. In this case, there are 19 operators anticommuting with $P$, all of which are listed in the top-left column of TABLE~\ref{table_circuit_1}. The products $PY_0X_aX_bX_c$ are listed in the top-middle column. We then select the first of these operators, $Y_0Z_1Y_2Y_3Z_4Z_5Z_6Y_7$, and multiply the other 18 by it, producing 18 new operators that contain only identities and $X$'s, as shown in the top-right column of TABLE~\ref{table_circuit_1}.
}

\begin{table}[]
\begin{tabular}{ccccc}
Pool operators &  & Gradient observables &  & Redefined operators \\
\\
+YXXX\underline{\hspace{2.3mm}}\underline{\hspace{2.3mm}}\underline{\hspace{2.3mm}}\underline{\hspace{2.3mm}} &  & - YZYYZZZY &  & - YZYYZZZY       \\
+YXX\underline{\hspace{2.3mm}}\underline{\hspace{2.3mm}}X\underline{\hspace{2.3mm}}\underline{\hspace{2.3mm}} &  & - YZYZZYZY &  & +\underline{\hspace{2.3mm}}\underline{\hspace{2.3mm}}\underline{\hspace{2.3mm}}X\underline{\hspace{2.3mm}}X\underline{\hspace{2.3mm}}\underline{\hspace{2.3mm}} \\
+YXX\underline{\hspace{2.3mm}}\underline{\hspace{2.3mm}}\underline{\hspace{2.3mm}}\underline{\hspace{2.3mm}}X &  & +YZYZZZZZ &  & +\underline{\hspace{2.3mm}}\underline{\hspace{2.3mm}}\underline{\hspace{2.3mm}}X\underline{\hspace{2.3mm}}\underline{\hspace{2.3mm}}\underline{\hspace{2.3mm}}X \\
+YX\underline{\hspace{2.3mm}}XX\underline{\hspace{2.3mm}}\underline{\hspace{2.3mm}}\underline{\hspace{2.3mm}} &  & - YZZYYZZY &  & +\underline{\hspace{2.3mm}}\underline{\hspace{2.3mm}}X\underline{\hspace{2.3mm}}X\underline{\hspace{2.3mm}}\underline{\hspace{2.3mm}}\underline{\hspace{2.3mm}} \\
+YX\underline{\hspace{2.3mm}}X\underline{\hspace{2.3mm}}\underline{\hspace{2.3mm}}X\underline{\hspace{2.3mm}} &  & - YZZYZZYY &  & +\underline{\hspace{2.3mm}}\underline{\hspace{2.3mm}}X\underline{\hspace{2.3mm}}\underline{\hspace{2.3mm}}\underline{\hspace{2.3mm}}X\underline{\hspace{2.3mm}} \\
+YX\underline{\hspace{2.3mm}}\underline{\hspace{2.3mm}}XX\underline{\hspace{2.3mm}}\underline{\hspace{2.3mm}} &  & - YZZZYYZY &  & +\underline{\hspace{2.3mm}}\underline{\hspace{2.3mm}}XXXX\underline{\hspace{2.3mm}}\underline{\hspace{2.3mm}}   \\
+YX\underline{\hspace{2.3mm}}\underline{\hspace{2.3mm}}X\underline{\hspace{2.3mm}}\underline{\hspace{2.3mm}}X &  & +YZZZYZZZ &  & +\underline{\hspace{2.3mm}}\underline{\hspace{2.3mm}}XXX\underline{\hspace{2.3mm}}\underline{\hspace{2.3mm}}X   \\
+YX\underline{\hspace{2.3mm}}\underline{\hspace{2.3mm}}\underline{\hspace{2.3mm}}XX\underline{\hspace{2.3mm}} &  & - YZZZZYYY &  & +\underline{\hspace{2.3mm}}\underline{\hspace{2.3mm}}XX\underline{\hspace{2.3mm}}XX\underline{\hspace{2.3mm}}   \\
+YX\underline{\hspace{2.3mm}}\underline{\hspace{2.3mm}}\underline{\hspace{2.3mm}}\underline{\hspace{2.3mm}}XX &  & +YZZZZZYZ &  & +\underline{\hspace{2.3mm}}\underline{\hspace{2.3mm}}XX\underline{\hspace{2.3mm}}\underline{\hspace{2.3mm}}XX   \\
+Y\underline{\hspace{2.3mm}}XX\underline{\hspace{2.3mm}}X\underline{\hspace{2.3mm}}\underline{\hspace{2.3mm}} &  & +YYYYZYZY &  & +\underline{\hspace{2.3mm}}X\underline{\hspace{2.3mm}}\underline{\hspace{2.3mm}}\underline{\hspace{2.3mm}}X\underline{\hspace{2.3mm}}\underline{\hspace{2.3mm}} \\
+Y\underline{\hspace{2.3mm}}XX\underline{\hspace{2.3mm}}\underline{\hspace{2.3mm}}\underline{\hspace{2.3mm}}X &  & - YYYYZZZZ &  & +\underline{\hspace{2.3mm}}X\underline{\hspace{2.3mm}}\underline{\hspace{2.3mm}}\underline{\hspace{2.3mm}}\underline{\hspace{2.3mm}}\underline{\hspace{2.3mm}}X \\
+Y\underline{\hspace{2.3mm}}X\underline{\hspace{2.3mm}}X\underline{\hspace{2.3mm}}X\underline{\hspace{2.3mm}} &  & +YYYZYZYY &  & +\underline{\hspace{2.3mm}}X\underline{\hspace{2.3mm}}XX\underline{\hspace{2.3mm}}X\underline{\hspace{2.3mm}}   \\
+Y\underline{\hspace{2.3mm}}X\underline{\hspace{2.3mm}}\underline{\hspace{2.3mm}}X\underline{\hspace{2.3mm}}X &  & - YYYZZYZZ &  & +\underline{\hspace{2.3mm}}X\underline{\hspace{2.3mm}}X\underline{\hspace{2.3mm}}X\underline{\hspace{2.3mm}}X   \\
+Y\underline{\hspace{2.3mm}}\underline{\hspace{2.3mm}}XXX\underline{\hspace{2.3mm}}\underline{\hspace{2.3mm}} &  & +YYZYYYZY &  & +\underline{\hspace{2.3mm}}XX\underline{\hspace{2.3mm}}XX\underline{\hspace{2.3mm}}\underline{\hspace{2.3mm}}   \\
+Y\underline{\hspace{2.3mm}}\underline{\hspace{2.3mm}}XX\underline{\hspace{2.3mm}}\underline{\hspace{2.3mm}}X &  & - YYZYYZZZ &  & +\underline{\hspace{2.3mm}}XX\underline{\hspace{2.3mm}}X\underline{\hspace{2.3mm}}\underline{\hspace{2.3mm}}X   \\
+Y\underline{\hspace{2.3mm}}\underline{\hspace{2.3mm}}X\underline{\hspace{2.3mm}}XX\underline{\hspace{2.3mm}} &  & +YYZYZYYY &  & +\underline{\hspace{2.3mm}}XX\underline{\hspace{2.3mm}}\underline{\hspace{2.3mm}}XX\underline{\hspace{2.3mm}}   \\
+Y\underline{\hspace{2.3mm}}\underline{\hspace{2.3mm}}X\underline{\hspace{2.3mm}}\underline{\hspace{2.3mm}}XX &  & - YYZYZZYZ &  & +\underline{\hspace{2.3mm}}XX\underline{\hspace{2.3mm}}\underline{\hspace{2.3mm}}\underline{\hspace{2.3mm}}XX   \\
+Y\underline{\hspace{2.3mm}}\underline{\hspace{2.3mm}}\underline{\hspace{2.3mm}}XX\underline{\hspace{2.3mm}}X &  & - YYZZYYZZ &  & +\underline{\hspace{2.3mm}}XXXXX\underline{\hspace{2.3mm}}X     \\
+Y\underline{\hspace{2.3mm}}\underline{\hspace{2.3mm}}\underline{\hspace{2.3mm}}\underline{\hspace{2.3mm}}XXX &  & - YYZZZYYZ &  & +\underline{\hspace{2.3mm}}XXX\underline{\hspace{2.3mm}}XXX    \\

\\
\\
Generators &  & CNOT - decoupled &  & Locally diagonalized \\
\\
- YZYYZZZY       &  & - Y\underline{\hspace{2.3mm}}XXXXYZ       &  & +Z\hspace{0.3mm}\underline{\hspace{2.3mm}}Z\hspace{0.3mm}Z\hspace{0.3mm}Z\hspace{0.3mm}Z\hspace{0.3mm}Z\hspace{0.3mm}Z       \\
+\underline{\hspace{2.3mm}}X\underline{\hspace{2.3mm}}X\underline{\hspace{2.3mm}}\underline{\hspace{2.3mm}}\underline{\hspace{2.3mm}}\underline{\hspace{2.3mm}} &  & +\underline{\hspace{2.3mm}}X\underline{\hspace{2.3mm}}\underline{\hspace{2.3mm}}\underline{\hspace{2.3mm}}\underline{\hspace{2.3mm}}\underline{\hspace{2.3mm}}\underline{\hspace{2.3mm}} &  & +\underline{\hspace{2.3mm}}Z\underline{\hspace{2.3mm}}\underline{\hspace{2.3mm}}\underline{\hspace{2.3mm}}\underline{\hspace{2.3mm}}\underline{\hspace{2.3mm}}\underline{\hspace{2.3mm}} \\
+\underline{\hspace{2.3mm}}\underline{\hspace{2.3mm}}X\underline{\hspace{2.3mm}}X\underline{\hspace{2.3mm}}\underline{\hspace{2.3mm}}\underline{\hspace{2.3mm}} &  & +\underline{\hspace{2.3mm}}\underline{\hspace{2.3mm}}X\underline{\hspace{2.3mm}}\underline{\hspace{2.3mm}}\underline{\hspace{2.3mm}}\underline{\hspace{2.3mm}}\underline{\hspace{2.3mm}} &  & +\underline{\hspace{2.3mm}}\underline{\hspace{2.3mm}}Z\underline{\hspace{2.3mm}}\underline{\hspace{2.3mm}}\underline{\hspace{2.3mm}}\underline{\hspace{2.3mm}}\underline{\hspace{2.3mm}} \\
+\underline{\hspace{2.3mm}}\underline{\hspace{2.3mm}}\underline{\hspace{2.3mm}}X\underline{\hspace{2.3mm}}X\underline{\hspace{2.3mm}}\underline{\hspace{2.3mm}} &  & +\underline{\hspace{2.3mm}}\underline{\hspace{2.3mm}}\underline{\hspace{2.3mm}}X\underline{\hspace{2.3mm}}\underline{\hspace{2.3mm}}\underline{\hspace{2.3mm}}\underline{\hspace{2.3mm}} &  & +\underline{\hspace{2.3mm}}\underline{\hspace{2.3mm}}\underline{\hspace{2.3mm}}Z\underline{\hspace{2.3mm}}\underline{\hspace{2.3mm}}\underline{\hspace{2.3mm}}\underline{\hspace{2.3mm}} \\
+\underline{\hspace{2.3mm}}\underline{\hspace{2.3mm}}\underline{\hspace{2.3mm}}\underline{\hspace{2.3mm}}X\underline{\hspace{2.3mm}}X\underline{\hspace{2.3mm}} &  & +\underline{\hspace{2.3mm}}\underline{\hspace{2.3mm}}\underline{\hspace{2.3mm}}\underline{\hspace{2.3mm}}X\underline{\hspace{2.3mm}}\underline{\hspace{2.3mm}}\underline{\hspace{2.3mm}} &  & +\underline{\hspace{2.3mm}}\underline{\hspace{2.3mm}}\underline{\hspace{2.3mm}}\underline{\hspace{2.3mm}}Z\underline{\hspace{2.3mm}}\underline{\hspace{2.3mm}}\underline{\hspace{2.3mm}} \\
+\underline{\hspace{2.3mm}}\underline{\hspace{2.3mm}}\underline{\hspace{2.3mm}}\underline{\hspace{2.3mm}}\underline{\hspace{2.3mm}}X\underline{\hspace{2.3mm}}X &  & +\underline{\hspace{2.3mm}}\underline{\hspace{2.3mm}}\underline{\hspace{2.3mm}}\underline{\hspace{2.3mm}}\underline{\hspace{2.3mm}}X\underline{\hspace{2.3mm}}\underline{\hspace{2.3mm}} &  & +\underline{\hspace{2.3mm}}\underline{\hspace{2.3mm}}\underline{\hspace{2.3mm}}\underline{\hspace{2.3mm}}\underline{\hspace{2.3mm}}Z\underline{\hspace{2.3mm}}\underline{\hspace{2.3mm}}
\end{tabular}
\caption{\BLUE{The top-left column contains all $iY_0X_jX_kX_l$ $S_z$ parity-preserving pool operators defined on 8 qubits that fail to commute with $P=Y_1Z_2Z_3Z_4Z_5Z_6Y_7$. The top-middle column contains the gradient observables, which are the commutators of $P$ with the pool operators (up to a factor of $2$), and it constitutes our initial observable set. The redefined observable set (top right) is obtained by multiplying every operator (but the first) with the first. 
The generator set (bottom left) is obtained by multiplying $X$-operators from the redefined set and removing redundant entries. Qubits 1--5 are disentangled by gates CNOT$_{13}$, CNOT$_{24}$, CNOT$_{35}$, CNOT$_{46}$, and CNOT$_{57}$ resulting in the CNOT-decoupled generator set (bottom middle). A combination of Phase and Hadamard gates completes the diagonalization of the generators (bottom right).}}\label{table_circuit_1}
\end{table}

\BLUE{
By taking products of the 18 $X$-operators, we can dispose of redundant ones, and put the array in an ``upper triangular'' form (the anchor column may need to be swapped with the $0^{th}$), thus isolating a reduced set of independent generators. Note that in this example, we obtain 5 $X$-operator generators and 6 generators in total, consistent with the fact that there can exist at most $2^N$ mutually commuting Pauli strings and by extension at most $N$ independent generators.
}

\BLUE{Next, we notice that the first operator is the only one that has a non-identity on the anchor qubit and can be diagonalized with single-qubit gates at that position. The same is not true for qubits for which two or more of the operators have distinct non-identity Paulis, as such qubits are entangled with others. The disentangling step involves applying a CNOT gate for each $X$-pair in the generator set, as $XX$ $\xrightarrow[]{\text{CNOT}}$ $XI$. The final step entails applying S and H gates to conclude the diagonalization. High-level pseudo-code for the entire procedure can be found in Appendix~\ref{appendix_algorithms}, Algorithm~\ref{alg:diagonalize_tableau}.}

\subsection{\label{subsec:shot_noise}Shot-noise robustness and resource estimation}

In subsection~\ref{subsec:singlepivot} we saw that it is possible to arrange the commutator of the Hamiltonian and \BLUE{all qubit pool operators into $\mathcal{O}(N^5)$} commuting sets.
However, the number of commuting sets is not by itself enough to determine the number of state preparations required for the task.

\BLUE{In the na\"ive approach where each Pauli $C_{ij}$ in the gradient operator $[H,A_i]$ is measured independently in separate experiments, the expected} error $\varepsilon_i$ in the estimate of this gradient is given by
\begin{align}\label{eq:variance}
    \varepsilon_i^2=\textrm{Var}(\langle[H,A_i]\rangle)=\sum_j \lvert c_{ij} \rvert^2 \frac{\textrm{Var}( C_{ij})}{s_{ij}},
\end{align}
where $s_{ij}$ is the number of independent samples of observable $C_{ij}$, and $\textrm{Var}( C_{ij})=1-\langle C_{ij}\rangle^2\in[0,1]$. We can then minimize $\varepsilon_i^2$ under total shot budget $S_{ni}=\sum_js_{ij}$ by defining the Lagrangian\footnote{The subscript $n$ is just a label for the na\"ive approach.} 
\begin{equation}
    \mathcal{L}_n=\sum_j \lvert c_{ij} \rvert^2 \frac{\textrm{Var}( C_{ij})}{s_{ij}}+\lambda_n\bigg(\sum_js_{ij}-S_{ni}\bigg).
\end{equation}
We find
\begin{align}\label{eq:shots_proportionality}
    s_{ij}=\frac{\lvert c_{ij} \rvert \sqrt{\textrm{Var}(C_{ij})}}{\sqrt{\lambda_n}},
\end{align}
with
\begin{align}\label{eq:lmultiplier}
    \frac{1}{\sqrt{\lambda_n}}=\frac{\sum_k\lvert c_{ik} \rvert \sqrt{\textrm{Var}(C_{ik})}}{\varepsilon_i^2}.
\end{align}
Setting all variances to their upper bound\footnote{This is a conservative but unphysical overestimate, for if all variances were equal to one, then all expectation values would vanish.}, we see that allotting each observable a number of samples directly proportional to its absolute weight is optimal:
\begin{equation}\label{eq:upper_bound_prop}
    s_{ij} = \frac{\lvert c_{ij} \rvert}{\varepsilon_i^2} \bigg( \sum_k \lvert c_{ik} \rvert \bigg)
\end{equation}
\BLUE{with
\begin{equation}
    S_{ni}=\frac{\big(\sum_k\lvert c_{ik} \rvert\big) ^2}{\varepsilon_i^2}.
\end{equation}
}

Now consider the case where we \BLUE{simultaneously measure groups of commuting Hamiltonian terms}:
\begin{equation}
    [H,A_i]=\sum_{g=1}^{\gamma_i} G_{ig} = \sum_{g=1}^{\gamma_i}\sum_{j=1}^{\zeta_{ig}}c_{igj}C_{igj},
\end{equation}
where $\gamma_i$ is the number of commuting sets, and $\zeta_{ig}$ is the number of observables in set $g$.
\BLUE{Ignoring covariance effects and} performing a similar minimization as before, we obtain for the expected number of state preparations for the measurement of \BLUE{the gradient of operator $A_i$}:
\begin{equation}\label{eq:shots_per_gradient}
    S_{gi}=\frac{ \bigg(\sum_{g=1}^{\gamma_i}\sqrt{\sum_{j=1}^{\zeta_{ig}}\lvert c_{igj}\rvert^2}\bigg)^2}{\varepsilon_i^2}.
\end{equation}

\BLUE{In our proposal, each of the simultaneously measured observables enters a different gradient, thus Eq.~\eqref{eq:variance} holds. However, Eqs.~\eqref{eq:shots_proportionality},~\eqref{eq:lmultiplier} consider each gradient separately for shot allocation. We can use global information by minimizing the sum of the variances of the gradients being simultaneously measured, $\sum_i \varepsilon^2_i$, where $i$ now ranges over all $\mathcal{O}(N^3)$ pool operators in one of the $2N$ commuting sets. Similarly to Eq. (\ref{eq:gradient_commutator}), we write each variance, Eq. (\ref{eq:variance}), such that for each term $h_jH_j$ in the Hamiltonian there exists a term $\lvert c_{ij}\rvert ^2 \frac{\textrm{Var}(C_{ij})}{s_{ij}}$ in each variance, perhaps with vanishing $\lvert c_{ij}\rvert$. Doing so allows us to combine terms from distinct gradients (different $i$ indices) arising from the same Hamiltonian term (sharing $j$ indices) which have equal denominators $s_{ij}\rightarrow s_j$, and switch the summation order:}
\BLUE{
\begin{equation}\label{eq:Lagrangian_proposal}
    \mathcal{L}_p=\sum_{j}  \frac{\sum_i \lvert c_{ij} \rvert^2 \textrm{ Var}( C_{ij})}{s_{j}}+\lambda_p\bigg(\sum_js_{j}-S_p\bigg),
\end{equation}
finding that the optimal number of shots for the $j^\mathrm{th}$ commuting set is
\begin{equation}\label{eq:shots_per_group}
    s_j=\frac{1}{\sqrt{\lambda_p}}\sqrt{\sum_i \lvert c_{ij} \rvert^2 \textrm{ Var}(C_{ij})}
\end{equation}
with
\begin{equation}\label{eq:multiplier_proposal}
    \frac{1}{\sqrt{\lambda_p}}=\frac{\sum_k\sqrt{\sum_i\lvert c_{ik} \rvert^2\textrm{Var}(C_{ik})}}{\sum_i\varepsilon_i^2}.
\end{equation}
Summing over Hamiltonian terms and once again setting all variances to their upper bound, we obtain for one set of commuting pool operators:}

\BLUE{
\begin{equation}\label{eq:total_shots_for_comm_set}
    S_p= \frac{\bigg( \sum_k \sqrt{\sum_i\lvert c_{ik} \rvert^2}\bigg) ^2}{\sum_i \varepsilon_i^2}.
\end{equation}}

\BLUE{We can use the fact that for nonvanishing commutators $\lvert c_{ik} \rvert =2 \lvert h_k \rvert$, set each gradient's expected error to $\varepsilon$, and make the simplifying assumption that each Hamiltonian term anticommutes with half the pool operators in the commuting set to obtain}
\BLUE{
\begin{equation}\label{eq:estimate_for_comm_set}
    \Tilde{S}_p= \frac{ 2~\big( \sum_k \lvert h_{k} \rvert \big) ^2}{ \varepsilon^2},
\end{equation}
or twice as many shots as na\"ively measuring the energy once. This implies that measuring all gradients to precision $\varepsilon$ is expected to be $4N$ times as expensive as a na\"ive VQE iteration.
}

\BLUE{Eqs.~\eqref{eq:total_shots_for_comm_set} and~\eqref{eq:shots_per_gradient} look strikingly similar. However, we note that the sum under the square root in Eq.~\eqref{eq:shots_per_gradient} is over commuting terms of the same gradient each corresponding to a different Hamiltonian term (thus, having unequal weights in general), whereas the sum under the root in Eq.~\eqref{eq:total_shots_for_comm_set} is over pool operators anticommuting with the same Hamiltonian term (and having equal coefficients). We see that in either case dividing high-weight observables into many sets instead of combining them in a few is undesirable, from Eqs.~\eqref{eq:shots_per_gradient}, \eqref{eq:total_shots_for_comm_set}} and the properties of the square root. To wit:
\begin{align*}
    \sqrt{1^2+2^2}+\sqrt{1^2+2^2} > \sqrt{1^2+1^2}+\sqrt{2^2+2^2}.
\end{align*}
Thus, we should prioritize grouping large-coefficient observables together, so that shots are not squandered over-measuring unimportant terms. In the case of our proposed strategy, where for nonvanishing commutators $\lvert c_{ij} \rvert \propto \lvert h_j \rvert$, we see that this requirement is saturated: \textit{All} commutators in each set have identical coefficients.

 Not only does the proposed grouping of gradient observables result in significantly fewer commuting sets, but it does so in a manner conscious of measurement statistics. For these reasons, we expect our Hamiltonian term-by-term partitioning to be resilient to finite sampling error and impactful in terms of shot economy, especially considering that molecular Hamiltonian coefficients typically span at least 10 orders of magnitude even for small molecules. We note that the heuristic algorithms for combining Hamiltonian terms into commuting sets for energy measurement purposes in Refs. \cite{Crawford2021efficientquantum} and \cite{ghost_paulis} prioritize combining large-coefficient observables with very promising results.

\subsection{\BLUE{Simulations}}\label{subsec:simulations}

\BLUE{In this subsection, we numerically compare our proposal for measuring gradients using pool operator grouping against standard gradient-by-gradient measurement strategies (i.e., no grouping of pool operators). The latter is done in two ways: (i) by na\"ively measuring all terms comprising a single operator gradient individually, and (ii) by simultaneously measuring mutually  commuting groups of terms comprising each gradient, with the partitioning done using the shot-noise aware Sorted Insertion heuristic. In Sorted Insertion, the Pauli terms of an operator are sorted by descending absolute coefficient. Then, each term is placed into the first set with which it commutes with all members. If no such set exists, a new one is created~\cite{Crawford2021efficientquantum}. We also provide na\"ive energy measurement shot counts and compare to the analytic estimate presented in Subsec.~\ref{subsec:shot_noise}. Results are presented for three molecular systems: H$_2$, H$_4$, and H$_6$ which are 4-, 8-, and 12-qubit problems, respectively, in the STO-3G basis set. Due to the classical computational cost of these simulations, we opted for using a reduced operator pool for each system such that the potential of each of the three methods is fairly evaluated. Specifically, we measured the gradients of the mutually commuting set of $\{iY_0X_jX_kX_l\}$ operators, which amounts to $\frac{1}{2N}$ of the full qubit pool for each system. For H$_2$ the set has only one element: $iY_0X_1X_2X_3$, for H$_4$ it has 19, and for H$_6$ it has 85 elements. Our simulation code makes use of the Qiskit~\cite{qiskit2024}, OpenFermion~\cite{McClean_2020}, and stim~\cite{gidney2021stim} Python modules which handle statevector simulations, Pauli observables, and diagonalization circuits, respectively. At each run, Qiskit takes as inputs an ADAPT ansatz, a measurement circuit and the number of shots, and outputs a list of bit-strings and their frequencies of occurrence. The shot allocation is dictated by Eqs.~\eqref{eq:variance}-\eqref{eq:multiplier_proposal} with covariances accounted for where necessary. Variances and covariances are sequentially estimated by measuring for each energy or gradient, all Pauli observables $r$ times as many shots as Eqs.~\eqref{eq:variance}-\eqref{eq:multiplier_proposal} dictate, $\forall ~r \in R$. In our simulations $R=(0.01, 0.4, 1)$ for H$_2$ and $R=(0.4, 1)$ for H$_4$ and H$_6$, with initial guess variances set to 1 and guess covariances to 0. Precision $\varepsilon$ was set to $10^{-3}$ for all measured gradients and energies.
}
\\
\begin{figure*}[ht]
\centering   
\includegraphics[width=1\linewidth]{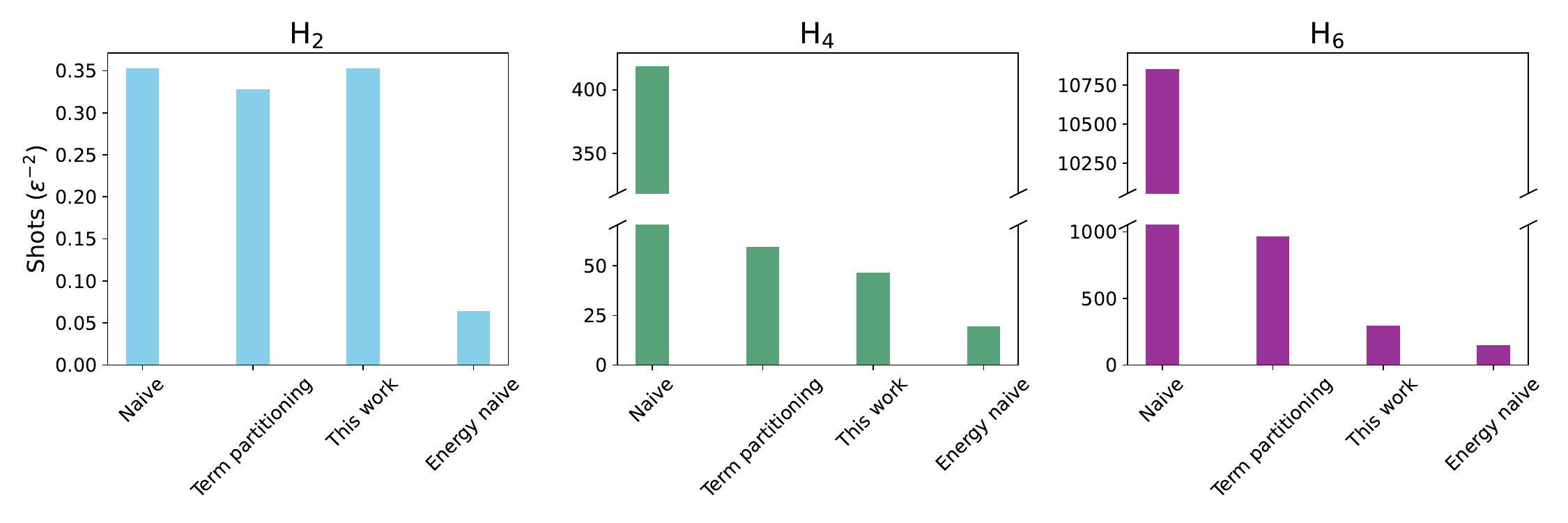}
  \caption{Shot counts for the na\"ive approach to gradient measurement, measurement gradient-by-gradient with term partitioning, measurement using our pool operator grouping strategy, and na\"ive energy measurement for H$_2$, H$_4$, and H$_6$.}
  \label{fig:moneymaker1}
\end{figure*}

\BLUE{
Our main results are presented in FIG.~\ref{fig:moneymaker1}, with the corresponding arithmetic values in TABLE~\ref{table:shots}. In the leftmost panel, we see that for the almost trivial H$_2$ problem, our proposal and the na\"ive approach coincide since there is only one gradient to be measured. Simultaneously measuring terms of the gradient partitioned into 2 sets only has a slight advantage over the other two methods because of relatively large positive covariances in this system and specific term partitioning.}

\BLUE{Doubling the system size, we already see from the middle panel of FIG.~\ref{fig:moneymaker1} that both our proposed strategy and the more standard gradient-by-gradient measurement with term partitioning yield an order of magnitude improvement compared to na\"ively measuring the gradients. H$_4$ is a great probe into the argument presented in Subsec. \ref{subsec:shot_noise}. The Sorted insertion heuristic \cite{Crawford2021efficientquantum} that we use for term partitioning prioritizes grouping large-coefficient terms together, and for H$_4$ it divides all 19 gradients' terms into only 182 groups, while the Hamiltonian has 184 terms other than the identity. Even though the two methods require measuring roughly the same number of sets of observables, we see that simultaneously measuring observables with unequal weights (measuring gradient by gradient) requires 28\% more shots compared to measuring sets of observables with equal absolute coefficients (our strategy) to reach accuracy~$\varepsilon$ for all gradients. Compared to na\"ively measuring gradients, our method achieves an 89\% shot reduction.} 

\BLUE{Increasing the problem size by another 4 qubits widens the gap between our proposal and measuring gradient by gradient. For H$_6$, we see from the right panel of FIG.~\ref{fig:moneymaker1} that our scheme cuts the number of state preparations by 69\% compared to measurement with term partitioning and by 97\% compared to na\"ive measurement, reflecting the $\mathcal{O}(N^3)$ speedup.}

\begin{figure}
  \includegraphics[width=1\linewidth]{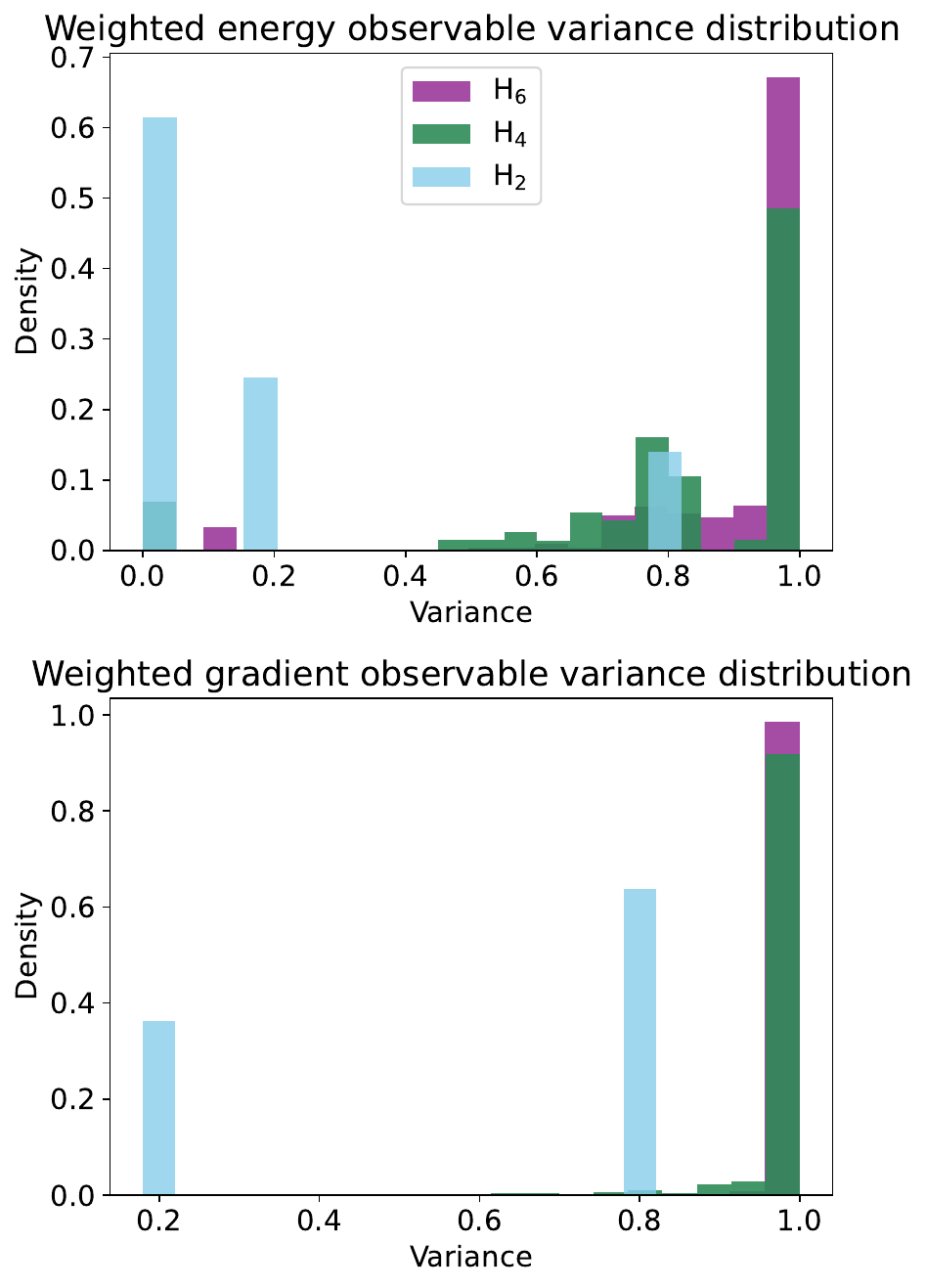}
  \caption{Normalized distributions of the exact variances of Pauli observables required for energy (top) and gradient (bottom) measurement for H$_2$, H$_4$, and H$_6$, weighted by the absolute coefficient of each Pauli.}
  \label{fig:variances}
\end{figure}

\BLUE{In Subsec.~\ref{subsec:shot_noise} we argued that the expected number of shots needed for measuring the gradients of a commuting set of pool operators using our method is twice the number of shots required for measuring the energy na\"ively. For H$_2$, gradient measurement is almost three times more expensive than our estimate, and for H$_4$ it exceeds our estimate by 19\%. To derive Eq.~\ref{eq:estimate_for_comm_set}, we set each Pauli observable variance to its maximum value of 1 and assumed that half of the terms in the inner sum of Eq.~\eqref{eq:total_shots_for_comm_set} vanish, on average. Clearly, either or both assumptions break down for smaller systems. To understand the reason, we turn to Fig.~\ref{fig:variances}, where we plot the normalized distributions of variances of the Pauli observables required for energy and gradient evaluation for the three systems, weighted by absolute coefficients. We see that our maximum variance assumption holds better for gradient observables than for energy observables. This is because molecular Hamiltonians contain $\mathcal{O}(N^2)$ diagonal terms (e.g. $Z_0, Z_1, ... , Z_0Z_1, Z_0Z_2, ... $) which often have large expectation values and smaller variances. For this reason, we expect our estimate to be more accurate for larger system sizes as the $\mathcal{O}(N^4)$ two-body terms dominate the Hamiltonian, which can also be seen in the top panel of Fig.~\ref{fig:variances} as we move from H$_2$ to H$_6$. Indeed, for H$_6$, our analytic estimate is correct to within about 0.5\% of simulation. We briefly discuss and numerically test the second assumption in Appendix ~\ref{appendix_assumptions}}.

\begin{table}[]
\begin{tabular}{lrrrr}
\hline
     & \multicolumn{1}{c}{Na\"ive} & \multicolumn{1}{c}{Term partitioning} & \multicolumn{1}{c}{This work} & \multicolumn{1}{c}{Energy na\"ive} \\ \hline
H$_2$ & 352776                                     & 327596                               & 353229                        & 64086                                            \\
H$_4$ & 418685970                                  & 59426817                             & 46454367                      & 19505730                                          \\
H$_6$ & 10855668777                                & 967276862                            & 298315571                     & 148246274                                         \\ \hline
\end{tabular}
\caption{Shot counts for the simulations discussed in Subsec.~\ref{subsec:simulations}, corresponding to the results in Fig.~\ref{fig:moneymaker1}. }
\label{table:shots}
\end{table}

\begin{figure}
  \includegraphics[width=1\linewidth]{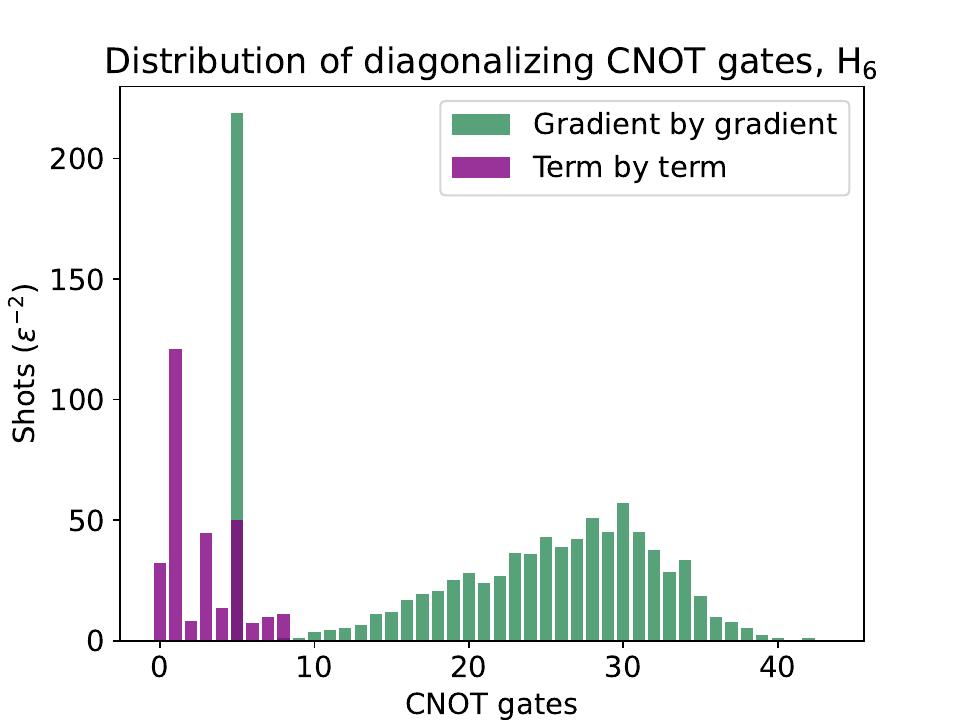}
  \caption{Distribution of diagonalizing circuit CNOT gates for gradient measurement term-by-term with pool operator grouping (our proposal) and gradient-by-gradient with term grouping. Each circuit is weighted by the number of times it was executed.}
  \label{fig:cnots}
\end{figure}

 \BLUE{In the spirit of using hardware-efficient operator pools for ADAPT-VQE, we aim to curb the diagonalizing CNOT overhead associated with simultaneous measurement. In Fig.~\ref{fig:cnots}, we juxtapose our highly structured diagonalization circuit problem to using a successful generic-use diagonalization procedure applied to the less-structured problem of diagonalizing commuting terms of the same gradient. Specifically, for H$_6$, we look at the distribution of CNOT gates required for diagonalization in our proposed measurement scheme with circuits as presented in Subsec.~\ref{subsec:diagonalizing_circ}, and compare it to measurement by gradient term partitioning and circuits compiled using the method introduced in Ref. \cite{vandenBerg2020circuitoptimization}, and Algorithm 3 therein. Fig.~\ref{fig:cnots} demonstrates that even for a small system, our deterministic partitioning offers another important advantage against a more general way to combine commuting observables. The fundamental reason for this is that in our proposal we simultaneously measure commutators of a pivot with many \textit{qubit-wise} commuting gradients, while in the case of gradient-by-gradient measurement with term partitioning we measure commutators of a pivot with many generally-commuting (i.e., not necessarily qubit-wise) Hamiltonian terms, which generally involves diagonalizing circuits with at least $\Omega(\frac{N^2}{\mathrm{log}N})$ CNOT gates.} 

\section{Discussion}\label{sec:discussion}

\BLUE{In this work, we proposed an efficient gradient-measurement strategy for the top-performing hardware-efficient operator pools in ADAPT-VQE based on the simultaneous measurement of commuting observables. We exploit the structure of the problem to arrange the $\mathcal{O}(N^8)$ Paulis into $\mathcal{O}(N^5)$ sets of mutually commuting observables. We find that the $\mathcal{O}(N^3)$ Paulis in each set can be diagonalized with circuits containing at most $N-3$ CNOT gates, and the fact that all elements of any one set have equal weights in the gradients allows for a quasi-optimal allocation of shots. We show that measuring the gradients of the entire pool is only $4N$ times as expensive as na\"ively measuring the expectation value of the Hamiltonian, to the same accuracy.}

\BLUE{In our approach, measuring the $\mathcal{O}(N^4)$ gradients can be thought of as measuring the energy na\"ively (i.e., term-by-term of the Hamiltonian), $2N$ times, and as such it is incompatible with advancements made for energy measurement based on Hamiltonian term grouping. However, approaches that reduce the energy measurement overhead of VQE by reducing the 1-norm of the Hamiltonian, $\sum_j\left|h_j\right|$, are directly applicable to ours, as gradient measurement is directly proportional to the square of the 1-norm, as discussed in Subsec.~\ref{subsec:shot_noise}. Notable methods include transformations to localized orbital bases~\cite{orbital_transformations}, and Hamiltonian factorizations~\cite{RCDF_hohenstein}.}

\BLUE{Although we focus on the qubit, QEB, and CEO pools with molecular Hamiltonians, our work is applicable to other systems and pools, without necessarily providing an advantage. For example, in ADAPT-QAOA~\cite{ADAPT_QAOA}, the Max-Cut problem is mapped onto an Ising-type Hamiltonian which decomposes into $\mathcal{O}(N^2)$ $ZZ$-type terms, and a $\mathcal{O}(N^2)$-sized operator pool of $XX$-, $YY$-, and $YZ$-type operators. It is easy to see that the Hamiltonian terms all commute, whereas the pool operators can be partitioned into $N+2$ commuting sets. That is, for large, densely connected graphs, measuring term-by-term of the Hamiltonian as in our strategy implies measuring $\mathcal{O}(N^3)$ sets of observables, whereas measuring gradient-by-gradient would be preferable since the number of sets is as large as the operator pool, and scales as $\mathcal{O}(N^2)$.}

\BLUE{Throughout our manuscript we assumed we have access to noiseless state preparation, which is far from true for current and near-future devices. Dalton and colleagues have shown~\cite{Quantifying_gate_errors}, by simulating ansatz preparation under depolarizing noise for a range of molecules, that ADAPT-VQE with hardware-efficient operator pools has the best chance of reaching chemical accuracy, owing to its compact circuits. In particular, they found that the maximum allowed gate-error probability to reach chemical accuracy is roughly inversely proportional to the number of noisy CNOT gates. We believe that ADAPT-VQE experiments on NISQ devices using our measurement scheme, which only incurs a minimal CNOT overhead, perhaps coupled with 1-norm reduction techniques, is a natural future research direction.}

\section{Conclusions}\label{sec:conclusions}
\BLUE{
In this work we introduced a partitioning of $\mathcal{O}(N^4)$ Pauli observables into $2N$ mutually commuting sets by proving and leveraging the fact that the commutators of a set of commuting Pauli words with any one Pauli word mutually commute. We argued that our proposed partitioning automatically considers shot-noise effects, since by design all observables in any one set have the same absolute coefficients in the quantities of interest (the gradients), and each set can be allotted a quasi-optimal number of state preparations according to its importance.}

\BLUE{We showed that under reasonable assumptions the number of shots required to measure the pool gradients to error $\varepsilon$ is only $4N$ times as expensive as na\"ively measuring the expectation value of the Hamiltonian to the same precision.
We proved that simultaneous diagonalization circuits with at most $N-3$ CNOT gates exist, and we presented a straightforward procedure for finding them.
Our approach can significantly reduce the state preparation overhead of ADAPT-VQE, and we expect it to find use in implementations of ADAPT-VQE on real hardware.
}

\section*{Code and Data Availability}

All data was generated with the code publicly available at \url{https://github.com/panastasiou137/ADAPT-gradients/} and is available upon request.

\section*{Author contributions}
P.G.A.: Methodology, Software, Formal analysis, Visualization, Writing - Original draft.
N.J.M.: Conceptualization, Manuscript - Review \& Editing
E.B.: Conceptualization, Methodology, Formal analysis, Validation, Manuscript - Review \& Editing.
S.E.E.: Conceptualization, Methodology, Validation, Supervision, Manuscript - Review \& Editing.

\section*{acknowledgments}

We thank Yanzhu Chen for helpful discussions.
S.E.E. and P.G.A. were supported by the U.S. Department of Energy, Office of Science, National Quantum Information Science Research Centers, Co-design Center for Quantum Advantage (C2QA) under contract number DE-SC0012704. 
E.B. and N.J.M. acknowledge support from Award No. DE-SC0019199 and DE-SC0025430.

\providecommand{\noopsort}[1]{}\providecommand{\singleletter}[1]{#1}%
%


\appendix

\section{Operator Pools}\label{appendix_operator_pools}

Of special interest to the quantum simulation community is the electronic Hamiltonian, which in second quantization can be written as
\begin{equation}\label{electronic_hamiltonian}
    H=\sum_{p,q} h^p_{q}a_p^{\dagger}a_q+\frac{1}{2}\sum_{p,q,r,s} h^{pq}_{sr}a_p^{\dagger}a_q^{\dagger}a_ra_s,
\end{equation}
where $h^p_q$ and $h^{pq}_{sr}$ are one- and two-electron integrals, respectively. In the original implementation of ADAPT-VQE, molecular Hamiltonians as in Eq.~\eqref{electronic_hamiltonian} were studied using pools comprised of UCC-type antihermitian sums of single and double fermionic excitation operators, $\hat{T}_{ij}=a_i^{\dagger}a_j-a_j^{\dagger}a_i$ and $\hat{T}_{ijkl}=a_i^{\dagger}a_j^{\dagger}a_ka_l-a_l^{\dagger}a_k^{\dagger}a_ja_i$; it was shown that ADAPT with this pool significantly outperforms the Trotterized UCCSD ansatz in both accuracy and compactness \cite{OG_ADAPT}. Under the JW transform, which we use in the remainder of this work, for $i<j<k<l$ these operators become
\begin{align}\label{single_excitation}
    \hat{T}_{ij}&=\frac{i}{2}(X_iY_j-Y_iX_j)\prod_{p=i+1}^{j-1}Z_p, 
\end{align}
and
\begin{align}
    \hat{T}_{ijkl}&=\frac{i}{8}(X_iY_jX_kX_l+Y_iX_jX_kX_l+Y_iY_jY_kX_l \nonumber \\
    &+Y_iY_jX_kY_l-X_iX_jY_kX_l-X_iX_jX_kY_l \nonumber \\
    &-Y_iX_jY_kY_l-X_iY_jY_kY_l)\prod_{p=i+1}^{j-1}Z_p\prod_{p=k+1}^{l-1}Z_p. \label{double_excitation}
\end{align}
It is evident, however, that the rotations these \BLUE{operators} generate translate into circuits with many entangling gates and act on $\mathcal{O}(N)$ qubits each. In order to avoid the cumbersome circuits associated with fermionic pools, Tang et al. introduced qubit-ADAPT-VQE \cite{qubitADAPT}, in which the sums in Eqs.~\eqref{single_excitation} and \eqref{double_excitation} are broken up into individual terms, the trailing Pauli $Z$ strings are removed, and the resulting constant Pauli weight operators are added to the pool instead. These are, up to index permutations: $iY_iX_j$, $iY_iX_jX_kX_l$, $iX_iY_jY_kY_l$. Qubit-ADAPT-VQE achieves a significant decrease in the number of CNOT gates required for ansatz preparation compared to the original fermionic ADAPT-VQE, at the expense of additional iterations and variational parameters. Shortly after, Qubit-Excitation Based (QEB-)ADAPT-VQE was developed \cite{QEB_ADAPT_VQE}, in which the pool operators are single and double ``qubit excitations" (QE), which are equivalent to only omitting the Z strings in Eqs.~\eqref{single_excitation} and \eqref{double_excitation}. In conjunction with the efficient circuits introduced in the same work, the QE pool yields ans\"atze with at least as few CNOT gates as qubit-ADAPT, using significantly fewer iterations. 
\BLUE{The state-of-the-art in this line of research is the Coupled Exchange Operator (CEO) pool recently introduced by Ramôa and colleagues~\cite{Ramoa2025}, in which operators are sums and differences of QE operators such that only 4 of the 8 terms survive. CEO evolutions can be implemented by even shallower circuits while still conserving relevant symmetries.}

It can be easily shown using the fermionic anticommutation relations, that commutators $[a^\dagger_pa^\dagger_qa_ra_s,  a^\dagger_ia^\dagger_ja_ka_l]$ vanish for $\{i,j,k,l\}\cap\{p,q,r,s\}=\varnothing$, and can be written in terms of 2- and 3-RDM elements otherwise \cite{RDM-ADAPT}, which implies that the gradients of fermionic pools with up to double excitation operators can be measured in $\mathcal{O}(N^6)$ time. The same is not true, however, for the other three pools; in particular, qubit, QE, \BLUE{and CEO} pool operators can couple to fermionic Hamiltonian terms even when they do not share indices\footnote{This is perhaps easier to see in the qubit picture, where this happens whenever an odd number of qubit operator indices overlap with the parity Z-strings of one- and two-body Hamiltonian terms.}, and evaluating their gradients requires measuring $\mathcal{O}(N^8)$ observables. However, the success of these pools in producing accurate ans\"atze, while cutting down on the number of quantum operations \cite{qubitADAPT, QEB_ADAPT_VQE, Ramoa2025} and circuit depths \cite{tetris} makes them prime candidates for use on NISQ devices, and reducing the runtime and measurement cost of ADAPT-VQE with such pools is crucial.

\section{\label{appendix_Hlinearpartitions}Hamiltonian terms can be partitioned into \texorpdfstring{$\mathcal{O}(N^3)$}{O(N3)} mutually commuting sets.}

Perhaps the most obvious way to put the result of \ref{subsec:simultaneous_commutation_proof} to use is to form commuting sets of Hamiltonian terms and measure the gradient of each pool operator sequentially. In the JW mapping, Hamiltonian two-body terms (whose number dominates for large $N$) take the form
\begin{align} \label{eq:two-body-term}
a_p^{\dagger}a_q^{\dagger}a_ra_s+a_s^{\dagger}a_r^{\dagger}a_qa_p \longrightarrow\frac{1}{8}(Y_pY_qX_rX_s \nonumber \\ +X_pX_qY_rY_s-X_pX_qX_rX_s  -Y_pX_qY_rX_s \nonumber \\ -X_pY_qY_rX_s-Y_pX_qX_rY_s-X_pY_qX_rY_s\nonumber \\-Y_pY_qY_rY_s)\prod_{t=s+1}^{r-1}Z_t\prod_{u=q+1}^{p-1}Z_u
\end{align}
for $p > q > r > s$, where it can be easily seen that the 8 sub-terms commute. Furthermore, it is easy to show using the fermionic anticommutation relations that $[a_p^{\dagger}a_q^{\dagger}a_ra_s, a_i^{\dagger}a_j^{\dagger}a_ka_l]=0$ when the two two-body terms share no indices, i.e. $\{i,j,k,l\}\cap\{p,q,r,s\}=\varnothing$. It is empirically known that for chemical Hamiltonians, using graph-coloring approaches to arrange the $\mathcal{O}(N^4)$ terms into commuting sets yields $\mathcal{O}(N^3)$ commuting sets \cite{GC_toQWC_Unitaries_Izmaylov}, and it was more recently shown \cite{Pranav_ON3} that for $N$ divisible by 4, a partitioning of the two-body terms into at most ${N-1 \choose 3} \sim N^3$ sets of disjoint-index terms always exists, by invoking Baranyai's theorem: $N$ vertices (qubit/spin-orbital indices) of a hypergraph can be partitioned into size-4 subsets (the 4 distinct indices in a two-body term) in ${N \choose 4} \frac{4}{N}$ ways such that each 4-element subset appears in one of the partitions exactly once \cite{Hamiltonian_hypergraphs}. The number of sets can in general be reduced to less than ${N \choose 4} \frac{4}{N}$, since not all 4-index combinations will appear in the Hamiltonian, and not all Hamiltonian terms will couple to the operator whose gradient we wish to measure; only about half. Thus, resorting to one of the previously proposed heuristic algorithms \cite{Crawford2021efficientquantum, ghost_paulis, min_clique_Izmaylov} is desirable. In any case, the $\mathcal{O}(N^3)$ asymptotic scaling of the number of commuting sets seems inescapable, and measuring the qubit pool gradients in this fashion would imply measuring the expectation values of observables arranged in $\mathcal{O}(N^7)$ sets in order to measure the gradients of the entire pool.

\section{Assumptions}\label{appendix_assumptions}

\BLUE{In order to derive our central result, Eq.~\eqref{eq:estimate_for_comm_set} we made the simplifying assumption that each Hamiltonian term anticommutes with about half of the operators in a commuting set. While it is true that the probability for any two random N-qubit Pauli strings to commute tends to $50\%$, the strings at hand are far from random.}

\begin{figure}[ht]
  \includegraphics[width=1\linewidth]{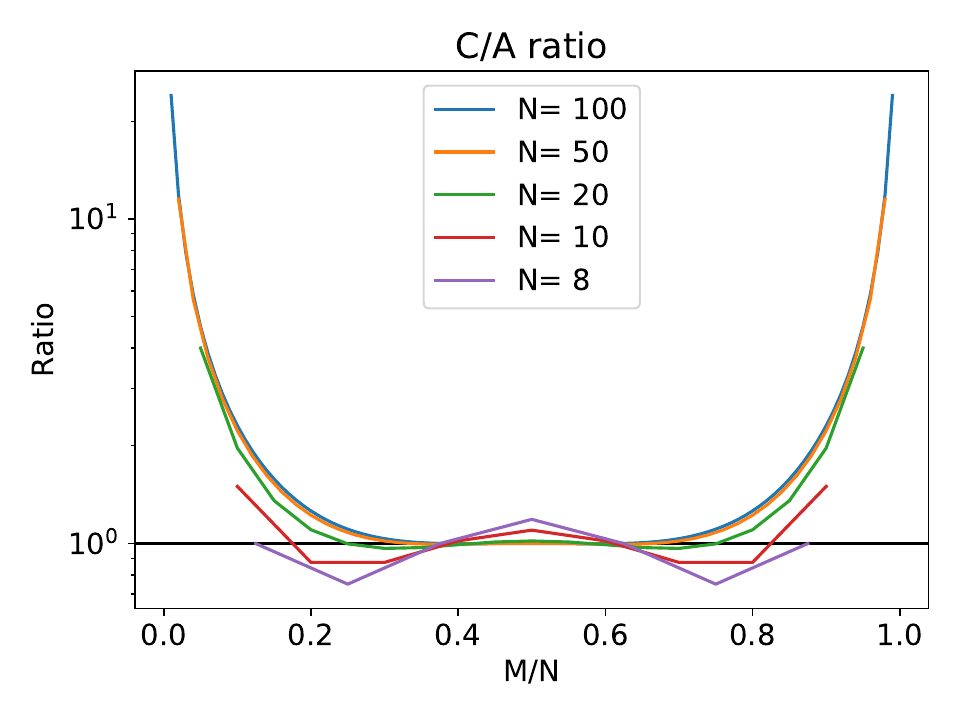}
  \caption{$\frac{C}{A}$ as a function of $\frac{M}{N}$, for a range of $N$.}
  \label{fig:ac_ratio}
\end{figure}

\BLUE{We begin by noticing that for larger systems, the parity $Z$-strings dominate the two-body terms (Eq. \ref{eq:two-body-term}) in the Hamiltonian, while for 1, 2 and 3 out of 4 indices $p, q, r, s$ coinciding with indices $i, j, k, l$ only half the terms in Eq. \ref{eq:two-body-term} anticommute with operators $Y_iX_jX_kX_l$ and $X_iY_jY_kY_l$. Thus, we focus on the $Z$-strings, and compute that an $M$-local $Z$-string, commutes with 
$C=8\big[\binom{M}{0}\binom{N-M}{4}+\binom{M}{2}\binom{N-M}{2}+\binom{M}{4}\binom{N-M}{0}\big]$ and anticommutes with $A=8\big[\binom{M}{1}\binom{N-M}{3}+\binom{M}{3}\binom{N-M}{1}\big]$ operators of the form $YXXX$, $XYYY$. The 
$\binom{M}{4}$ and $\binom{N-M}{4}$ terms dominate when $M \approx N$ and for highly local (such as the $\mathcal{O}(N^2)$ $Z_0, Z_1, Z_0Z_1...$) terms respectively, whereas the $\binom{M}{2}\binom{N-M}{2}$ term takes over in the neighborhood of $M=\frac{N}{2}$, and the two terms in $A$ are greater in the regions around $M=\frac{N}{4}$ and $M=\frac{3N}{4}$. In Fig.~\ref{fig:ac_ratio} we plot $\frac{C}{A}$ as a function of $\frac{M}{N}$ for a range of $N$. Overall, the products of even and odd binomial coefficients in $A$ and $C$ balance out for a wide range of $\frac{M}{N}$, in which the bulk of the Hamiltonian terms lie, and the ratio is close to one. For example, for the symmetric case of $M=M-N=\frac{N}{2}$, the ratio becomes exactly $\frac{C}{A}=\frac{N^2-4N+6}{N^2-4N}$. We note that at either extreme of $\frac{M}{N}$, all but one products of binomials in $C$, $A$ decay rapidly or vanish, with either $\binom{M}{4}$ or $\binom{N-M}{4}$ surviving, and tipping the scale in favor of commutation, thanks to a minority of highly local and global Hamiltonian terms. }

\begin{figure*}[ht]
\centering   
\includegraphics[width=1\linewidth]{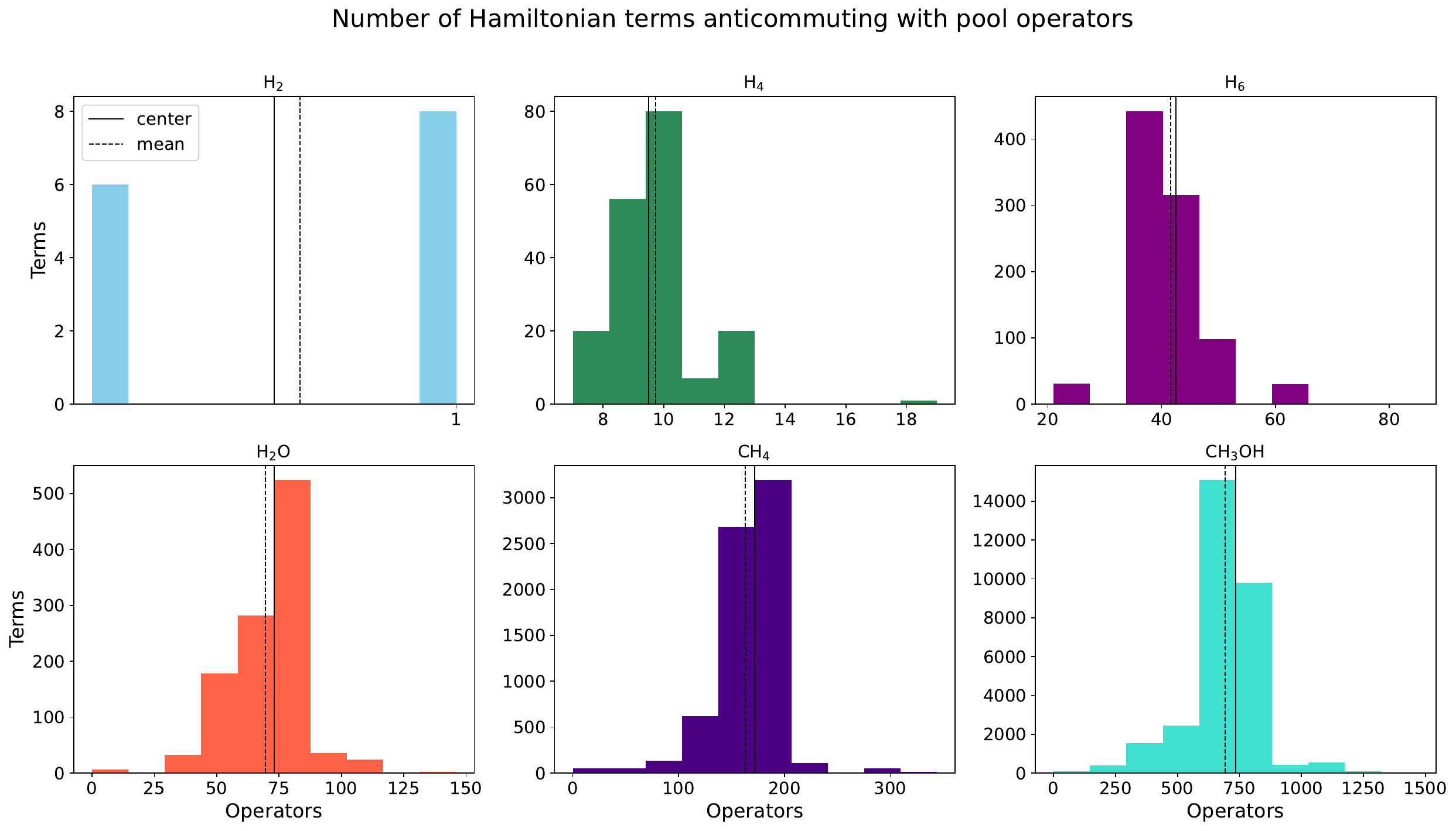}
  \caption{Distribution of Hamiltonian terms anticommuting with members of the $\{iY_0X_iX_jX_k\}$ sets of commuting operators for H$_2$, H$_4$, H$_6$, H$_2$O , CH$_4$, and CH$_3$OH which are 4, 8, 12, 14, 18, and 28 spin-orbital (and qubit) problems. For each system, the average number of operators the Hamiltonian terms anticommute with (mean), as well as one half of the number of elements in the $\{iY_0X_iX_jX_k\}$ set (center) are indicated.}
  \label{fig:anticommutation_dist}
\end{figure*}

\BLUE{We once again turn to simulation in order to verify our predictions. In Fig. \ref{fig:anticommutation_dist}, for a range of molecular systems ranging from 4 to 28 spin orbitals, we plot the distribution of Hamiltonian terms anticommuting with members of the $\{iY_0X_iX_jX_k\}$ sets of commuting operators as well as the average number of operators each Hamiltonian term anticommutes with. We see that for H$_2$ and H$_4$ the mean is slightly above one half of the number of elements of the $\{iY_0X_iX_jX_k\}$ set whereas for H$_6$ and larger systems it lies just below the mean, validating our assumption.}

\newpage

\newpage

\section{Algorithms}\label{appendix_algorithms}

\begin{algorithm}[H]
\caption{Generate commuting sets of gradient observables}\label{alg:generate_observables}
\begin{algorithmic}
\Require 
\Statex Qubit pool $\mathbf{Q}=\{iY_iX_j, iY_iX_jX_kX_l, iX_iY_jY_kY_l\}$. 
\Statex Hamiltonian $\mathbf{H}=\{(h_p, H_p)\}$ with $h_p$ complex coefficients and $H_p$ Pauli words. 
\Statex $N$ qubits.
\Ensure
\State $\mathbf{C} \gets \{\}$
\For {$H_p \in \mathbf{H}$}
    \For {$m = 0 \text{ to } N-1$}
        \State $\mathbf{K}_{\text{ypm}} \gets \{\}$
        \For {$Q_i=iY_mX_j, iY_mX_jX_kX_l \text{ in } \mathbf{Q}$}
            \If {$[Q_i, H_p]\neq0$}
                \State \text{add} $[Q_i, H_p]=c_{ip}C_{ip}$ \text{to} $\mathbf{K}_{\text{ypm}}$ \text{as} $(c_{ip}, C_{ip})$
            \EndIf
        \EndFor
        \State \text{add } $\mathbf{K}_{\text{ypm}}$ \text{ to } $\mathbf{C}$
        \State $\mathbf{K}_{\text{xpm}} \gets \{\}$
        \For {$Q_i=iX_mY_j, iX_mY_jY_kY_l \text{ in } \mathbf{Q}$}
            \If {$[Q_i, H_p]\neq0$}
            \State \text{add} $[Q_i, H_p]=c_{ip}C_{ip}$ \text{to} $\mathbf{K}_{\text{xpm}}$ \text{as} $(c_{ip}, C_{ip})$
            \EndIf
        \EndFor
        \State \text{add } $\mathbf{K}_{\text{xpm}}$ \text{ to } $\mathbf{C}$
    \EndFor
\EndFor
\State \Return $\mathbf{C}$
\end{algorithmic}
\end{algorithm}

\newpage

\begin{algorithm}[H] 
\caption{Generate diagonalizing circuit}\label{alg:diagonalize_tableau} 
\begin{algorithmic}
\Require 
\Statex List of $M$ commuting observables (stabilizers) generated by Algorithm~\ref{alg:generate_observables}. 
\Statex $N$ qubits. Anchor qubit $t$.

\State SWAP column $t$ with column 1 
\For{$m=2 \text{ to } M$} 
    \State MULTIPLY stabilizer $m$ by stabilizer 1 
\EndFor 

\For{$n=2 \text{ to } N$} 
    \State $s \gets \text{identify a stabilizer with nonidentity at column } n$ 
    \State SWAP stabilizer $s$ with stabilizer $n$ 
    \For{$m=n+1 \text{ to } M$} 
        \If{\text{stabilizers } $m, n$ \text{ coincide at column } $n$} 
            \State MULTIPLY stabilizer $m$ by stabilizer $n$ 
        \EndIf 
    \EndFor 
\EndFor 

\For{$m=M \text{ down to } 1$} 
    \For{$n=1 \text{ to } N$} 
        \For{$k=m-1 \text{ to } 1$} 
            \If{\text{stabilizers } $k, m$ \text{ coincide at column } $n$} 
                \State MULTIPLY stabilizer $k$ by stabilizer $m$ 
            \EndIf 
        \EndFor 
    \EndFor 
\EndFor 

\For{$m=1 \text{ to } M$} 
    \If{\text{stabilizer } $m$ \text{ is } $X_iX_j$} 
        \State APPLY gate $\text{CNOT}_{ij}$ 
    \EndIf 
    \If{\text{stabilizer } $m$ \text{ is } $Y_iY_j$} 
        \State APPLY gates $\text{CNOT}_{ij}\text{S}_j$ 
    \EndIf 
\EndFor 

\For{$n=1 \text{ to } N$} 
    \If{\text{every row has } $I$ \text{ OR } $Y$ \text{ at column } $n$} 
        \State APPLY gate $\text{S}_n$ 
    \EndIf 
    \If{\text{every row has } $I$ \text{ OR } $X$ \text{ at column } $n$} 
        \State APPLY gate $\text{H}_n$ 
    \EndIf 
\EndFor 

\Statex \Comment{\textit{Swapping columns $t$ and 1 can be accounted for classically by swapping gate targets $t$ and 1 everywhere in the circuit.}}
\end{algorithmic} 
\end{algorithm}

\end{document}